\documentclass[twocolumn,twoside,slac_two]{revtex4}
\usepackage{graphicx}
\usepackage{fancyhdr}
\begin{document}

\title{FACT - Monitoring Blazars at Very High Energies}

\newcommand{\uniw}{$^1$}
\newcommand{\ethz}{$^2$}
\newcommand{\unige}{$^3$}
\newcommand{\rwth}{$^4$}
\newcommand{\tudo}{$^5$}

\author{
D.~Dorner\uniw,
M.~L.~Ahnen\ethz,
M.~Bergmann\uniw,
A.~Biland\ethz,
M.~Balbo\unige,
T.~Bretz\ethz,
J.~Bu\ss\tudo,
S.~Einecke\tudo,
J.~Freiwald\tudo,
C.~Hempfling\uniw,
D.~Hildebrand\ethz,
G.~Hughes\ethz,
W.~Lustermann\ethz,
K.~Mannheim\uniw,
K.~Meier\uniw,
S.~Mueller\ethz,
D.~Neise\ethz,
A.~Neronov\unige,
A.-K.~Overkemping\tudo,
A.~Paravac\uniw,
F.~Pauss\ethz,
W.~Rhode\tudo,
T.~Steinbring\uniw,
F.~Temme\tudo,
J.~Thaele\tudo,
S.~Toscano\unige,
P.~Vogler\ethz,
R.~Walter\unige,
A.~Wilbert\uniw}

\affiliation{\uniw Universit\"at W\"urzburg, Institute for Theoretical Physics and Astrophysics, Emil-Fischer-Str.~31, 97074 W\"urzburg, Germany}
\affiliation{\ethz ETH Zurich, Institute for Particle Physics, Otto-Stern-Weg 5, 8093 Zurich, Switzerland}
\affiliation{\unige University of Geneva, ISDC Data Center for Astrophysics, Chemin d'Ecogia~16, 1290 Versoix, Switzerland}
\affiliation{\tudo TU Dortmund, Experimental Physics 5, Otto-Hahn-Str.~4, 44221 Dortmund, Germany}

\begin{abstract}

The First G-APD Cherenkov Telescope (FACT) was built on the Canary
Island of La Palma in October 2011 as a proof of principle for silicon
based photosensors in Cherenkov Astronomy. The scientific goal of the
project is to study the variability of active galatic nuclei (AGN) at
TeV energies. Observing a small sample of TeV blazars whenever
possible, an unbiased data sample is collected. This allows to study
the variability of the selected objects on timescales from hours to
years. Results from the first three years of monitoring will be
presented.

To provide quick flare alerts to the community and trigger
multi-wavelength observations, a quick look analysis has been installed
on-site providing results publicly online within the same night. In
summer 2014, several flare alerts were issued. Results of the quick
look analysis are summarized.

\end{abstract}

\maketitle

\thispagestyle{fancy}


\section{First G-APD Cherenkov Telescope}
The First G-APD Cherenkov Telescope (FACT) has been operational since October
2011 on the Canary Island of La Palma. It is located at 2200 meter
a.s.l.\ at the Observatorio del Roque de los Muchachos next to the
MAGIC Telescopes. Like its two neighbouring telescopes, FACT is a
ground-based gamma-ray telescope using the Imaging Air Cherenkov
Technique (IACT).  
\begin{figure}[t]
\centering
\includegraphics[width=80mm]{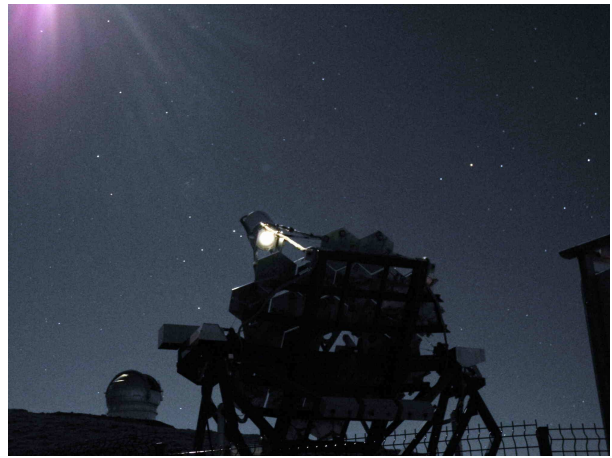}
\caption{First G-APD Cherenkov Telescope during the observations of the
full moon in June 2013. Credit: Daniela Dorner} \label{fact}
\end{figure}
The project uses the refurbished mount of HEGRA CT3 and the recoated
mirrors of HEGRA CT1, providing a mirror area of 9.5\,$m^2$. Figure
\ref{fact} shows the telescope during the observations of the full moon
in June 2013. In addition, FACT was equipped with a new drive system
and a novel camera using silicon based photosensors (SiPMs, a.k.a.\
Geigermode Avalanche Photo Diodes (G-APDs)). The camera consists of
1440\,pixels and has a total field of view of 4.5 degree. Details on
the design and construction of the telescope and the camera can be
found in \cite{fact-design}.

\subsection{Major Goals}
The technological challenge of the project was the proof of principle
of SiPMs in Cherenkov astronomy. So far, only photo-multiplier tubes
(PMTs) have been used in Cherenkov telescopes. To overcome the
limitations for observations during strong moonlight due to the aging
of the PMTs, different photosensors were considered. SiPMs were
selected, as they are very robust and do not show aging when exposed to
bright light. FACT is the first project which built a Cherenkov camera
equipped with SiPMs and tested it in regular operation. The camera has
now been operational since October 2011, and no SiPM has failed or
showed any problem or aging. 

The scientific goal of the FACT operations is the long-term monitoring
of bright TeV blazars. At very high energies (VHE), the available
observation time is limited, as IACTs run in pointing mode and have a
limited field of view. Active galactic nuclei (AGN), however, are
highly variable objects. As the measured variablity time scales of
blazars range from minutes to years, long-term monitoring is
mandatory. Although the shock-in-jet scenario can explain some features
of the variability, the picture is still inconclusive lacking
continuous data sets sampling a wide range of time scales. The
observation schedule of FACT focusses on a small sample of bright
blazars with the goal to obtain an unbiased and even sampling over the
entire visibility period for the sources.

\subsection{Profiting from SiPMs}
Several aspects of SiPMs make them the ideal photosensors for a
monitoring instrument. First of all, the robustness towards bright
ambient light allows to extend the observations into the full moon
period. It has been shown that even when observing the full moon,
showers can still be recorded \cite{icrc-moon}. Observing during strong
moonlight makes it possible to reduce the gaps in the light curves. 

It is known that the gain of SiPMs depends on the temperature and the
applied voltage and with the latter on the ambient light. As both the
temperature and the ambient light are changing, a feedback system was
introduced to keep the gain of the SiPMs stable. Details on the studies
of the gain stablity and on the feedback system can be found in
\cite{fact-feedback}. 

Keeping the gain stable and homogeneous over the camera, no calibration
of the photosensors is needed. For an independent crosscheck of the
calibration, FACT is equipped with an external light pulser. 

Studying the dependence of the trigger threshold on the ambient light,
it is possible to set the trigger threshold directly according to the
measured currents \cite{fact-feedback}. Knowing the position of the
moon and the observed sources, it is also possible to predict the
threshold for each observation \cite{icrc-threshold}. In this way, a
constistent performance of the system is achieved which facilitates the
analysis of the data. Furthermore, the stable performance allows to use
the rate of background events or other parameters for data quality
checks as shown in \cite{icrc-monitoring} and \cite{fact-feedback}.

\subsection{Towards Robotic Operation}
\begin{figure}[t]
\centering
\includegraphics[width=70mm]{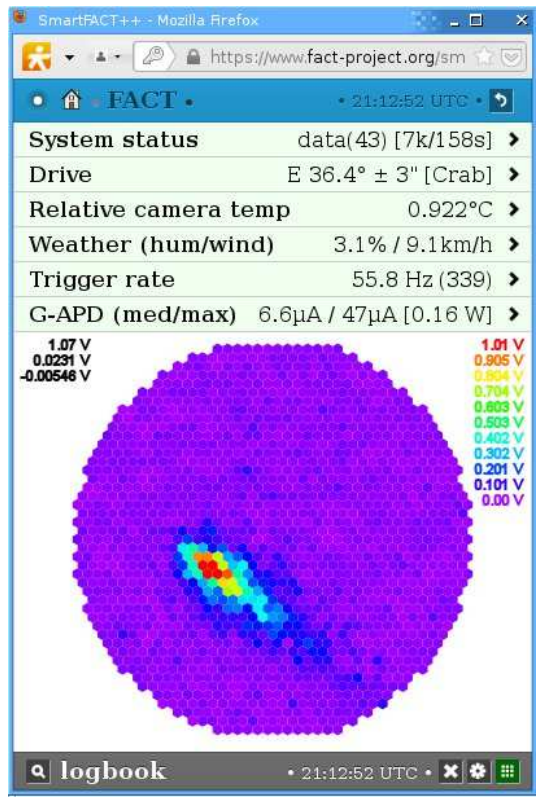}
\caption{Smartfact: a web interface to monitor the system and operate
the telescope.} \label{smartfact}
\end{figure}
The stable performance of the detector also allows for remote and
automatic operation of the telescope. While in the first few months,
there was a shift crew on-site, very soon remote operations started. In
addition, the operation was automatized step by step. In the meantime,
the system can be operated via a web interface which has been optimized
for smartphone-usage. In Figure \ref{smartfact}, a screenshot of the web
interface can be seen. It allows not only to monitor the instrument and
all its subsystems, but after authentication also to control the
system. The control has been automatized such that in the meantime the
shifter just needs to start a script at the beginning of the night,
monitor the system and the weather during the night and stop the script
at the end of the night. Known issues like communication problems are
automatically taken care off by the script. The shifter only needs to
interfere in case of bad weather or when an unknown problem appears. In
case of emergencies that need immediate intervention at the system, an
agreement ensures support from the MAGIC shift crew.

\section{Long-term Monitoring}
\subsection{Scientific Motivation}
As blazars show variability on time scales from minutes to years,
long-term monitoring is mandatory to uncover the mechanisms driving the
variability due to the propagation of perturbations down the jet. Short
snapshots with more sensitive larger telescopes such as MAGIC are
necessary to understand the underlying physics of the central engine
encoded in the shortest variability time scales. Multi-wavelength
observations are very important to study the spectral aging of the
emission regions during their propagation down the jet and to reveal
the underlying radiation mechanisms. The spectral energy distribution
(SED) of these objects exhibits two bumps where the position of the
peaks depends on the type of the object. For high-frequency peaked
blazars, the high energy peak is located at GeV to TeV energies making
them the most interesting targets for FACT.

\subsection{Source Sample}
Being most sensitive at TeV energies, there is a limited number of
sources which are good candidates for the long-term monitoring program
of FACT. The two AGN most observed by FACT are Mrk\,421 and Mrk\,501
which were detected in 1992 and 1996 by Whipple \cite{mrk421,mrk501}
and have been observed since then in several multi-wavelength (MWL)
campaigns. However, in these MWL campaigns the observations are still
rather sparse, e.g.\ one observation every three to ten days. The two
sources were also part of the Whipple monitoring program where from
Mrk\,421 878.4 hours in 783 nights were collected in 14 years
\cite{mrk421-14years}. 

In the first three years of operation, FACT has observed Mrk\,421 for
537 hours during 291 nights and Mrk\,501 for 924 hours during 372
nights. Also, the sources 1ES\,2344+51.4, 1ES\,1959+650, 1ES\,1218+304,
IC\,310, 1H0323+342, RGB\,J0521.8+2112 and PKS\,0736+01 have been
observed. While the first two belong to the regular monitoring sample,
the other sources have been observed only for 10 to 60 hours. For
performance studies, also a lot of observations of the Crab Nebula have
been carried out. 

Including observations during strong moonlight, each year about 3000
hours of observations are available. Taking into account bad weather
and technical problems, about 1800 hours per year remain of which about
1500 hours are physics data.

\subsection{MWL Observations}
To draw conclusions on the underlying physics processes, it is
necessary to study both peaks of the SED. Because of the variability of
the sources, the observations must be carried out simultaneously. As
continuous monitoring is not available in most wavelength ranges, MWL
campaigns and Target of Opportunity (ToO) observations are appropriate
means to obtain reasonable temporal overlap. Due to its blazar
monitoring program and fast data pipeline, FACT can readily provide
suitable triggers for ToO observations.

\section{Quick Look Analysis}
To provide quick alerts for ToO observations, it is important to
analyse the data quickly. For this, a quick look analysis (QLA) was set
up on La Palma in December 2012. 

\subsection{Setup}
Once the data are written on disk, they are immediately processed by
the QLA. To avoid any interference with the data taking, the data are
transfered to another computer, where they are then analysed. The data
are typically recorded and processed in bunches of five minutes. For
technical reasons, also shorter durations of runs may occur. For the
QLA, the data are binned consequently in five minutes or multiples of
it. The data are processed with an automatic pipeline where each step
automatically starts the next one once it is finished. As a last step,
the results are inserted into a mysql database from where the plots for
a web interface are produced. 

\subsection{Open Access}
The results from the QLA are publicly available at {\tt
http://www.fact-project.org/monitoring} since September 2013. The data
are available in 20-minute and nightly binning. The results provided on
the website are excess rates. Details on the analysis and how the
excess rates are calculated follow in the next section. The results
shown on the website do not include any data quality check and also no
correction for the dependence of the threshold on the zenith distance
and ambient light. Therefore, the values are lower limits in case of
larger zenith distance or moonlight. Nevertheless, they are sufficient
to trigger other observations. Starting from March 2014, official flare
alerts to other instruments have been issued regularly.

\subsection{Analysis}
As the offline analysis in the data center, the QLA is also performed
with the Modular Analysis and Reconstruction Software (MARS)
\cite{mars}. Compared to the final analysis, the pipeline is optimized
to process the data as quickly as possible. 

In case a new analysis version is available, the old data are not
reprocessed with this new version. For example on May 26th 2014, a new
software version was introduced. For different software versions, the
results can differ slightly. 

The steps performed in the QLA, are shortly discussed in the following
paragraphs. For a quick estimate of the flux, it is not necessary to
calculate a spectrum using Monte Carlo simulations. Instead, the
excess-rate is used. 

\paragraph{Calibration} 
As mentioned before, calibrating the data is not necessary in case the
feeback system keeps the gain stable and homogeneous. When the QLA was
set up, the feedback system was not yet available in its final version.
Therefore, in the first version of the QLA the data were calibrated
using data from the external light pulsar. Starting from May 26th 2014,
this calibration is no longer used. 

In the step of the calibration, also the signal is extracted and bad
pixels are interpolated. 

\paragraph{Image Cleaning} 
Based on the extracted signal and extracted timing information, the
images are cleaned, i.e.\ pixels only containing noise are removed. For
this, two thresholds are applied for the signal and a time coincidence
window for the arrival time. First, all pixels with a signal higher
than threshold 1 are kept (core pixels). In addition, all neigbouring
pixels are kept in case their signal is higher than threshold 2.
Furthermore, pixels are only kept when their difference in arrival time
to the neighbouring pixel is smaller than 17.5\,ns. The thresholds for
the signal were first 4.0 and 2.5. For the new calibration, the levels
had to be increased to 5.2 and 3.3 to keep the excess rates roughly the
same. 

Next, a statistical analysis of the shower images is done, calculating
various image parameters which can be used to reconstruct the type,
the origin and the energy of the primary particle.

\paragraph{Background Suppression} 
In a first step, events that cannot be reconstructed are removed, i.e.\  events which 
\begin{itemize}
\item consist of only five pixels or less
\item have more than three islands
\item have $Leakage > 0.1$ where $Leakage$ is defined by the ratio of
signal in the outermost ring of pixels in the camera to the total
signal 
\end{itemize}
Next, the background suppression cuts are applied where the following
cuts are used: 
\begin{itemize}
\item $0.18 < SlopeSpreadWeighted < 0.68 $
\item $log10(Area)>(log10(Size)-2) \cdot 1.1-1.55$
\item $Conc_{core} < 0.13$
\item $Conc_{COG}  < 0.15$
\end{itemize}
where $Size$ and $Area$ the total amount of light in and the area
\cite{area} of the shower image. $Conc_{core}$ and $Conc_{COG}$ are the
ratio of the the signal in the core pixel and three pixels next to the
center of gravity to the total signal of the shower image, representing
the concentration of the light in the image. While $Slope$ is
development of the arrival time along a shower axis,
$SlopeSpreadWeighted$ is the spread of the slope along the main shower
axis weighted with the $Size$. 

For the reconstruction of the shower origin, the disp is calculated
\cite{disp}. The disp is parametrized as follows: 

$disp= \xi \cdot (1 - Width/Length) $ where $Width$ and $Length$ are
the standard deviations of the signal along the two shower axes. $\xi$
is the sum of the following correction terms: 
\begin{itemize}
\item constant term: $1.14136$
\item slope term: $ 0.0681437 \cdot Slope$
\item leakage term: $ 2.62932 \cdot log10(Leakage+1)$
\item size term: $ 0.0507821 \cdot (log10(Size)-1.51279)^2$ if $(log10(Size) > 1.51279)$
\end{itemize}

\paragraph{Excess-Rate Curves} 
\begin{figure}[t]
\centering
\includegraphics[width=80mm]{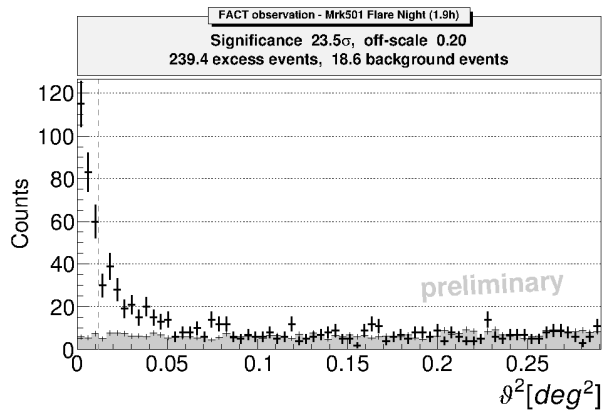}
\caption{$\theta^2$-plot of Mrk\,501 from 1.9 hours of data from the
night of 8th to 9th June 2012.} \label{thetasq}
\end{figure}
Next, $\theta^2$ is calculated as the distance between $Dist$ and
$disp$ where $Dist$ is the distance of the center of gravity to the
nominal source position. Plotting the $\theta^2$ distribution, one can
determine whether the source is detected. An example can be seen in
Figure \ref{thetasq}. The on-measurement is shown as black crosses, the
off measurement as gray area. The vertical dashed line represents the
cut in $\theta^2$ which determines the signal region. To calculate the
excess, the off-measurement is subtracted from the on-measurement in
the signal region. Deviding this by the ontime of the observation, the
excess rate is obtained which is a measure of the flux. As the trigger
threshold and therefore the energy threshold depend on the amount of
ambient light and the zenith distance, the same applies for the excess
rate. 

To study the variability of the sources in detail, either the excess
rate needs to be corrected for the effect of zenith distance and
ambient light or the flux needs to be calculated reconstructing the
energy with the help of simulated data. For sending alerts, the excess
rates are sufficient. Knowing the excess rates of the Crab Nebula, a
rough correction of the excess rates can be made. As a larger zenith
distance or more ambient light decrease the excess rate, the measured
excess rate can be considered as lower limit.

\section{Results}
The excess rates calculated in the QLA are used to get a rough idea of
the variability of the sources and to trigger ToO observations. Apart
from Mrk\,421 and Mrk\,501, the other sources did not show any
significant flares or activity in the last three years. In the
following, the excess rate curves of the former are discussed. To get
an overview of the activity, the excess rate curves are shown in a
nightly binning. As directly the results of the QLA are shown, no data
quality check or correction of the excess rate is included here. 

\subsection{Mrk\,421}
\begin{figure*}[t]
\centering
\includegraphics[width=170mm]{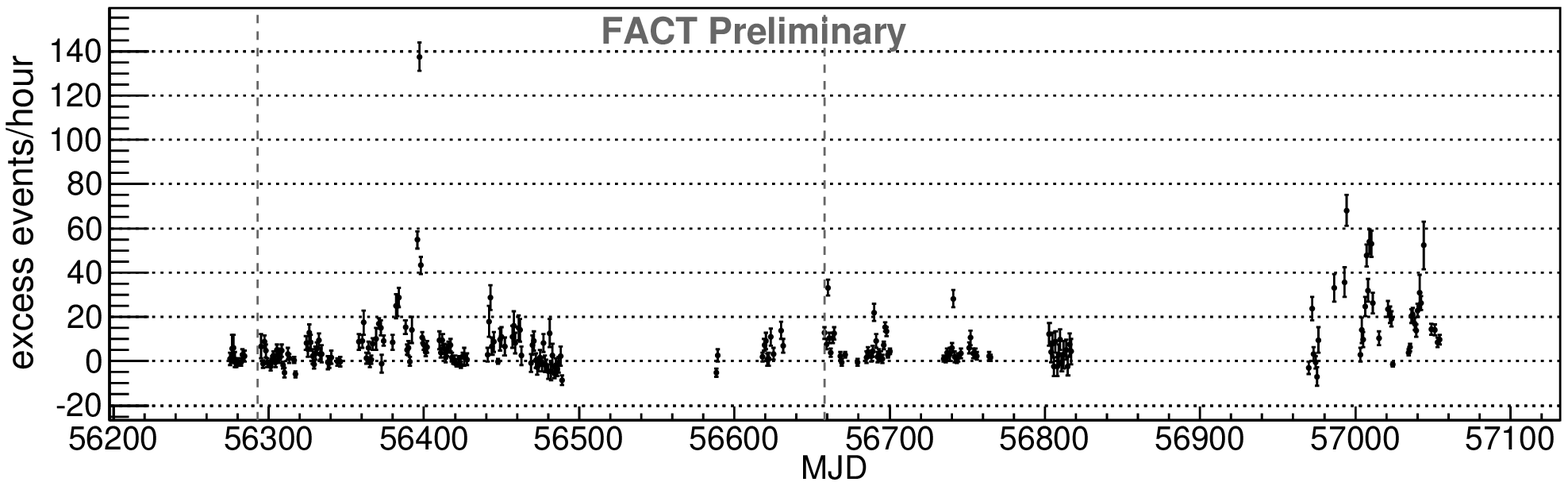}
\includegraphics[width=170mm]{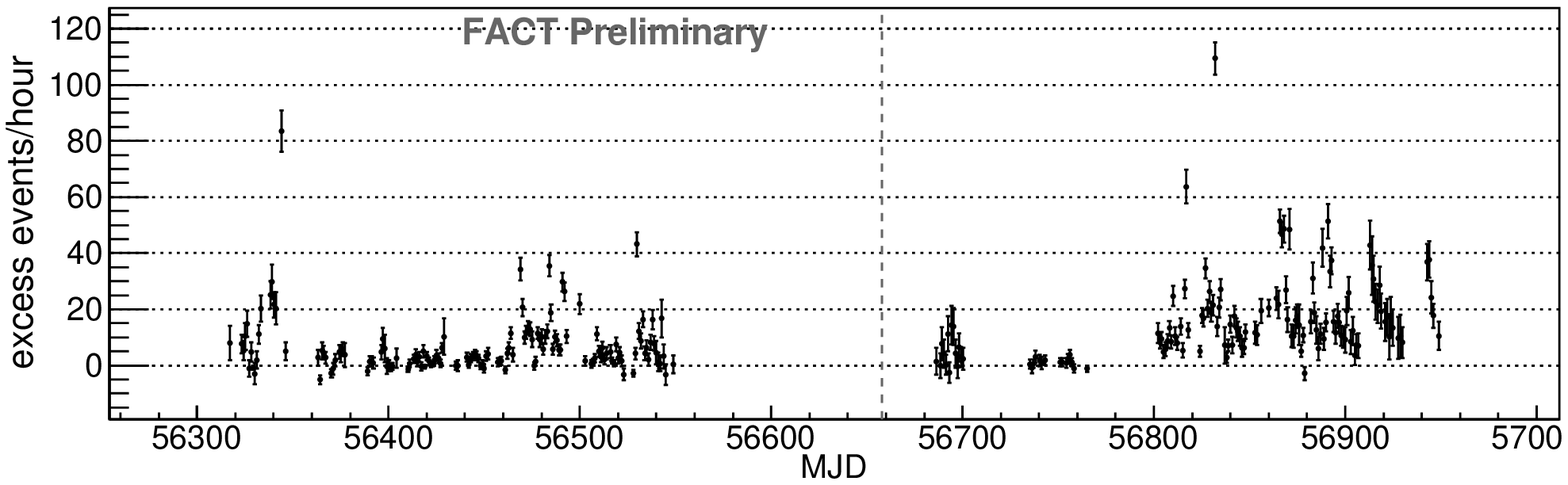}
\caption{Excess rate curves for Mrk\,421 (top panel) and Mrk\,501
(bottom panel) starting from December 12th 2012. The start of a new
year is marked with a gray, dashed, vertical line.} \label{excessrates}
\end{figure*}
Although Mrk\,421 was first observed in January 2012, here only the QLA
results starting from December 12th 2012 are shown. The excess rate
curve shown in Figure \ref{excessrates} in the top panel includes
roughly 500 hours of data. A big flare was observed in April 2013, and
in winter 2014/15 the source showed several times a flux higher than
the flux of the Crab Nebula. 

\subsection{Mrk\,501}
\begin{figure*}[t]
\centering
\includegraphics[width=170mm]{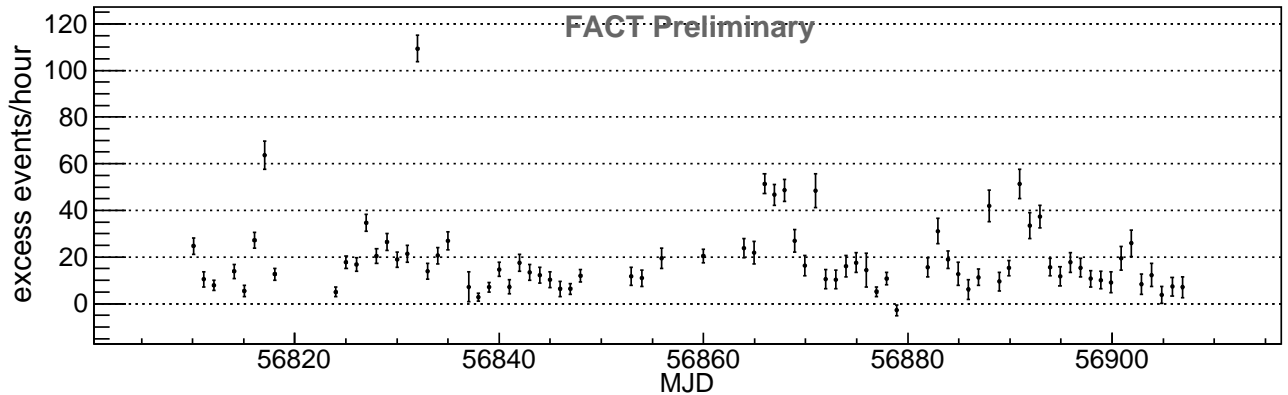}
\caption{Excess rate curve of Mrk\,501 from June to September 2014.}
\label{qlamrk501flare} \end{figure*}
Mrk\,501 was first observed in May 2012. In June 2012, a flare was
observed \cite{gamma2012,icrc-monitoring}. In Figure \ref{excessrates},
only the QLA results starting from December 12th 2012 are shown. The
excess rate curve shown here (lower panel) includes roughly 740 hours
of data. After a short flare in February 2013, Mrk\,501 showed some
moderate activity (around the flux of the Crab Nebula) in summer 2013.
In summer 2014, the source got more active with two flares and an
enhanced flux around 2-3 times the flux of the Crab Nebula.

\subsection{Flare Alerts}
In March 2013, FACT officially started sending alerts to other
instruments. Since then, seven flare alerts have been issued. Between
June and August 2014, six times an alert was sent as Mrk\,501 met the
trigger criteria of three Crab units. In Figure \ref{qlamrk501flare}, a
zoom in the excess rate curve to the time range of the flare alerts is
shown. As two alerts on 19th and 21st of June triggered observations of
other instruments, the big flare on June 24th was also observed by HESS
\cite{atelhess}.

\section{Summary and Outlook}
FACT using SiPMs has proven to be an ideal setup for the long-term
monitoring of blazars: The robustness of these photosensors allows to
extend the observations into times with strong moonlight closing the
gaps in light curves. Furthermore, the stable performance allows for a
high data taking efficiency and facilitates the analysis. 

To foster ToO observations, a quick look analysis was set up allowing
to send flare alerts within a short time. In 2014, seven flare alerts
have been issued. From Mrk\,501 and Mrk\,421, several flares and
periods with enhanced flux level have been observed. 

The next steps are to include in the quick look analysis an automatic
data quality check and a correction of the excess rates for the effect
that the trigger threshold changes with zenith distance and ambient
light.

\begin{acknowledgments}
The important contributions from ETH Zurich grants ETH-10.08-2 and
ETH-27.12-1 as well as the funding from the Swiss SNF, the German BMBF
(Verbundforschung Astro- und Astroteilchenphysik) and the DFG
(collaborative research center SFB 876/C3) are gratefully acknowledged.
We are thankful for the very valuable contributions from E. Lorenz, D.
Renker and G. Viertel during the early phase of the project. We also
thank the Instituto de Astrofisica de Canarias for allowing us to
operate the telescope at the Observatorio Roque de los Muchachos in La
Palma, the Max-Planck- Institut für Physik for providing us with the
mount of the former HEGRA CT 3 telescope, and the MAGIC Collaboration
for their support. We further thank the group of M. Tose from the
College of Engineering and Technology at Western Mindanao State
University, Philippines, for providing us with the scheduling web
interface.
\end{acknowledgments}


\end{document}